\documentclass[aps,twocolumn,showpacs,superscriptaddress]{revtex4-1}
\usepackage[T1]{fontenc}
\usepackage{amsfonts,amsmath,amssymb,amsbsy}
\usepackage{graphicx,color}
\usepackage{times}
\usepackage[colorlinks=true, linkcolor=blue, citecolor=red, urlcolor=magenta]{hyperref}

\let\mathbf=\boldsymbol
\def\D{{\mathcal{D}}}
\def\emph#1{\textcolor{red}{#1}}
\def\emph#1{\textcolor{black}{#1}}

\begin{document}

\title{Thermally stable magnetic skyrmions in multilayer synthetic antiferromagnetic racetracks}

\author{Xichao Zhang}
\affiliation{Department of Physics, University of Hong Kong, Hong Kong, China}
\affiliation{School of Electronic Science and Engineering, Nanjing University, Nanjing 210093, China}

\author{Motohiko Ezawa}
\email[]{ezawa@ap.t.u-tokyo.ac.jp}
\affiliation{Department of Applied Physics, University of Tokyo, Hongo 7-3-1, Tokyo 113-8656, Japan}

\author{Yan Zhou}
\email[]{yanzhou@hku.hk}
\affiliation{Department of Physics, University of Hong Kong, Hong Kong, China}
\affiliation{School of Electronic Science and Engineering, Nanjing University, Nanjing 210093, China}

\begin{abstract}
A magnetic skyrmion is a topological magnetization structure with a nanometric size and a well-defined swirling spin distribution, which is anticipated to be an essential building block for novel skyrmion-based device applications. We study the motion of magnetic skyrmions in multilayer synthetic antiferromagnetic (SAF) racetracks as well as in conventional monolayer ferromagnetic (FM) racetracks at finite temperature. There is an odd-even effect of the constituent FM layer number on the skyrmion Hall effect (SkHE). Namely, due to the suppression of the SkHE, the magnetic skyrmion has no transverse motion in multilayer SAF racetracks packed with even FM layers. It is shown that a moving magnetic skyrmion is stable even at room temperature ($T=300$ K) in a bilayer SAF racetrack but it is destructed at $T=100$ K in a monolayer FM racetrack. Our results indicate that the SAF structures are reliable and promising candidates for future applications in skyrmion-electronics and skyrmion-spintronics.
\end{abstract}

\date{\today}

\maketitle

\section{Introduction}
\label{se:Introduction}

A magnetic skyrmion is a nanometric magnetic domain wall structure of which the spin configuration is swirling in the planar space and would wrap a unit three-dimensional spherical surface with spins pointing in all directions in the compactification of the planar space~\cite{Roszler_NATURE2006,Ezawa_QHE2008,Nagaosa_NNANO2013,Bergmann_JPCM2014,Liu_CPB2015,Wiesendanger_Review2016}. Recently, experiments have revealed the existence of both N\'eel-type and Bloch-type skyrmions in thin films as well as bulk materials~\cite{Muhlbauer_SCIENCE2009,Yu_NATURE2010,Heinze_NPHYS2011,Seki_SCIENCE2012,Du_NANOLETT2014,Kezsmarki_NMATER2015,Nahas_NCOMMS2015,Woo_NMATER2016,Moreau-Luchaire_NNANO2016,Boulle_NNANO2016}. The spin configuration of a skyrmion is topologically protected, which is stable as long as it does not overlap tilted spins on the sample edge or shrink to the lattice scale. Another merit of a skyrmion is that it can be displaced by a spin-polarized current in confined structures. Thus, the skyrmion is expected to be a key component of future device applications in the emerging field of skyrmionics~\cite{Iwasaki_NC2012,Fert_NNANO2013,Sampaio_NNANO2013,Iwasaki_NNANO2013,Sun_PRL2013,Iwasaki_NL2014,Koshibae_NCOMMS2014,Tomasello_SREP2014,Yan_NCOMMS2014,Yan_NCOMMS2015,Xichao_SREP2015B,Koshibae_JJAP2015,Fusheng_NANOLETT2015,Upadhyaya_PRB2015,Beg_SREP2015,Crum_NCOMMS2015,Xichao_NCOMMS2016,Schutte_PRB2014}. Recent experiments have demonstrated the creation, manipulation, and elimination of skyrmions in confined geometries using cutting-edge technologies~\cite{Romming_SCIENCE2013,Finazzi_PRL2013,Du_NCOMMS2015,Nii_NCOMMS2015,Buttner_NPHYS2015,Wanjun_SCIENCE2015,Woo_NMATER2016,Moreau-Luchaire_NNANO2016,Boulle_NNANO2016}. However, one significant obstacle to the essential transmission process of skyrmions in devices is the skyrmion Hall effect (SkHE)~\cite{Zang_PRL2011,Wanjun_ARXIV2016}, where a skyrmion gains a transverse motion perpendicular to the imposed driving direction. As a consequence, the skyrmion, which carries encoded information, might be destructed when it passes a narrow racetrack-type channel~\cite{Parkin_SCIENCE2008,Parkin_NNANO2015} and/or when it travels at a high speed, by touching the edge of the device~\cite{Xichao_SREP2015A,Purnama_SREP2015,Xichao_NCOMMS2016}.

Recently we have shown in Ref.~\onlinecite{Xichao_NCOMMS2016} that a bilayer-skyrmion moves perfectly straight in the direction of the driving current in a bilayer synthetic antiferromagnetic (SAF) racetrack, where the SkHE is completely suppressed. The two skyrmions in the top and bottom ferromagnetic (FM) layers of the bilayer SAF racetrack have the opposite skyrmion numbers, and thus the potential transverse shift directions are opposite and canceled~\cite{Xichao_NCOMMS2016}. We note that the antiferromagnetic (AFM) skyrmions in monolayer~\cite{Xichao_SREP2016} and bilayer~\cite{Tretiakov_PRL2016} AFM racetracks are also free from the SkHE, where the AFM skyrmions have the skyrmion number of zero.

Another challenging problem is about the effect of thermal fluctuations, which is detrimental to the skyrmion-based device applications. A skyrmion is largely deformed and might be easily destroyed by the thermal fluctuations. Hence, the thermal effect is one of the important factors to realize the skyrmion-based device applications~\cite{Yu_NMATER2011,Yu_NCOMMS2012,Schulz_NPHYS2012,Oike_NPHYS2015}.

In this paper, we study the thermal fluctuation effect on the motion of skyrmions in multilayer SAF racetracks as well as in monolayer FM racetracks based on numerical simulations and a multilayer Thiele equation. The skyrmions in all FM layers of a multilayer SAF racetrack are bound to one single SAF multilayer skyrmion by the interlayer AFM exchange coupling. We call it a SAF $N$-layer skyrmion, where $N$ stands for the number of constituent FM layers.

First, we investigate the current-driven dynamics of a SAF $N$-layer skyrmion at zero temperature ($T=0$ K). We find the odd-even effect of the number $N$ of constituent FM layers on the SkHE, which is understood as the total skyrmion number of the skyrmions in each individual FM layer. Namely, the SAF $N$-layer skyrmion shows the SkHE only when $N$ is odd. Furthermore, we find that the velocity of the SAF $N$-layer skyrmion at a given driving current density is inversely proportional to $N$ when the driving current is applied only in the bottom FM layer. We also point out that when the driving current is applied in all FM layers, the velocity of the SAF $N$-layer skyrmion is basically equal to that of the FM monolayer skyrmion at a given driving current density.

\begin{figure*}[t]
\centerline{\includegraphics[width=1.00\textwidth]{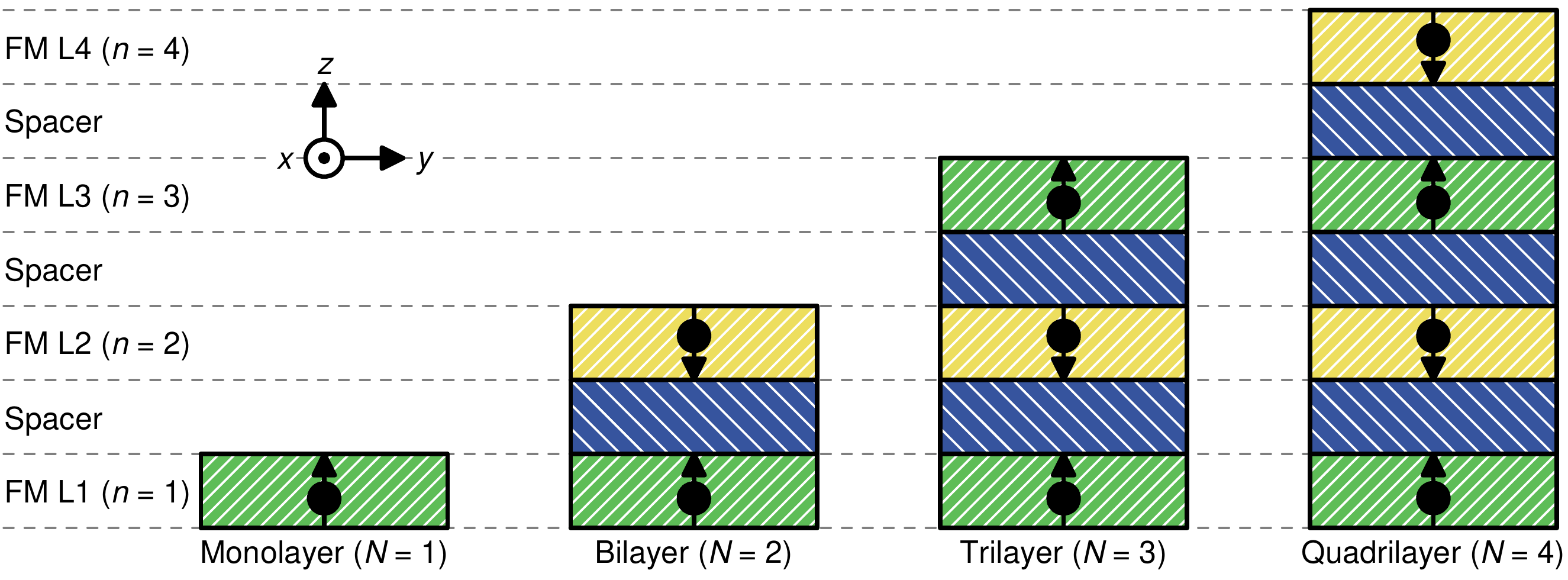}}
\caption{(Color online)
Schematics of the simulation models including the monolayer ($N=1$) FM, bilayer ($N=2$) SAF, trilayer ($N=3$) SAF and quadrilayer ($N=4$) SAF racetracks. The length along the $x$-axis, width along the $y$-axis, and thickness along the $z$-axis of each FM layer and nonmagnetic spacer are respectively equal to $500$ nm, $50$ nm and $1$ nm, where the FM layer L1 is placed on the heavy-metal substrate. In the SAF racetracks, the adjacent FM layers are antiferromagnetically exchange-coupled through their respective FM/spacer/FM interfaces. The initial magnetization configurations of FM layers L1 ($n=1$) and L3 ($n=3$) are pointing along the $+z$-direction, while those of FM layers L2 ($n=2$) and L4 ($n=4$) are pointing along the $-z$-direction. The arrows represent the initial magnetization directions.
}
\label{FIG1}
\end{figure*}

Second, we investigate the current-driven dynamics of a SAF $N$-layer skyrmion at finite temperature ($T>0$ K). It is shown that the FM monolayer skyrmion with typical material parameters is destroyed when $T=100$ K. On the other hand, the SAF bilayer skyrmion is stable and goes straight along the bilayer SAF racetrack even at room temperature ($T=300$ K). This is because that the two skyrmions consisting the SAF bilayer skyrmion are tightly bound by the interlayer AFM exchange coupling, where the SkHEs acting on the two skyrmions are opposite and canceled out. In general, for the system with a relatively large damping coefficient, the thermal fluctuation effect decreases rapidly as the layer number $N$ increases at a certain temperature. Our results provide a promising route to realizing the skyrmion-based device applications at room temperature.

\section{Modeling and Simulation}
\label{se:Modeling}

\subsection{Monolayer FM and multilayer SAF racetracks}
\label{se:Monolayer-and-multilayer}

Figure~\ref{FIG1} illustrates the monolayer FM racetrack as well as the multilayer SAF racetracks studied in this paper. The monolayer FM racetrack contains one FM layer and a heavy-metal substrate underneath the FM layer. The $N$-layer SAF racetrack ($N\geq2$) includes $N$ FM layers, which are separated by $N-1$ nonmagnetic spacer layers. The FM layers are denoted from bottom to top as L1, L2, L3, $\cdots$. In all models, the length along the $x$-axis, width along the $y$-axis, and thickness along the $z$-axis of each FM layer and nonmagnetic spacer are respectively equal to $500$ nm, $50$ nm, and $1$ nm, where the FM layer L1 is attached to the heavy-metal substrate.

In the $N$-layer SAF racetracks ($N\geq2$), the neighboring FM layers are antiferromagnetically exchange-coupled through their FM/spacer/FM interfaces. The magnetization in each FM layer is perpendicular to the racetrack plane due to the high perpendicular magnetic anisotropy (PMA), while the magnetization in neighboring FM layers are antiparallel due to the interlayer AFM exchange coupling. The Dzyaloshinskii-Moriya interaction (DMI) in FM layers lead to the tilt of magnetization near the edges of the FM layers. It is worth mentioning that the DMI in FM layers can be induced by both the heavy-metal substrate and spacer layers in real experiments~\cite{Wiesendanger_Review2016}. Recent theoretical~\cite{Yang_PRL2015,Dupe_NCOMMS2016} and experimental~\cite{Chen_APL2015,Stebliy_JAP2015,Moreau-Luchaire_NNANO2016,Boulle_NNANO2016,Woo_NMATER2016} studies have suggested and developed the methods to induce the DMI in multilayers. More promisingly, it has been recently shown that by constructing the spacer layer made of two different heavy-metal materials, additive DMI can be achieved in multilayers~\cite{Moreau-Luchaire_NNANO2016}.

In our numerical simulations we explicitly consider the SAF $N$-layer skyrmions with $N=1,2,3,4$. We assume that the relaxed magnetization distributions of the FM layers L1 and L3 are almost pointing along the $+z$-direction, while those of the FM layers L2 and L4 are almost pointing along the $-z$-direction. The background magnetization directions determine the skyrmion number of the skyrmion in each FM layer. The skyrmion number equals $1$ in the FM layers L1 and L3, while it equals $-1$ in the FM layers L2 and L4. The total skyrmion number $Q_{\text{tot}}$ of the SAF $N$-layer skyrmion is $N$ modulo $2$. Namely, $Q_{\text{tot}}=1,0,1,0$ for $N=1,2,3,4$, respectively (see Sec.~\ref{se:SkHE} for details).

At the initial state, the skyrmions are first created and relaxed at the position of $x=100$ nm, $y=25$ nm. With regard to the injection scheme of the driving current, we consider a confined current-perpendicular-to-plane (CPP) geometry. Namely, an electron current flows through the heavy-metal substrate in the $+x$-direction, which is converted into a spin current polarized along the $-y$-direction and is perpendicularly injected into the FM layer L1, due to the spin Hall effect (see Ref.~\onlinecite{Wanjun_SCIENCE2015} for a recent experimental example). The skyrmion in the FM layer L1 is driven by the vertical spin current, while the other skyrmions in the FM layers L2, L3, and L4 move accordingly due to the interlayer AFM exchange coupling between each adjacent FM layers. It should be noted that, for comparison purpose, we also simulate the straightforward case of unconfined CPP geometry with the bilayer SAF racetrack, where the spin current is injected into all FM layers, neglecting the spin current absorption in the model.

\subsection{Hamiltonian}
\label{se:Hamiltonian}

We investigate the multilayer SAF racetrack comprised of $N$ FM layers, where the neighboring FM layers are antiferromagnetically exchange-coupled by the interlayer AFM exchange interaction, as illustrated in Fig.~\ref{FIG1}. The total Hamiltonian $H$ is decomposed into the Hamiltonian for each FM layer $H_{n}$ and the interlayer AFM exchange coupling $H_{\text{inter}}$ between neighboring FM layers, that is,
\begin{equation}
H=\sum_{n=1}^{N}H_{n}+H_{\text{inter}}.
\label{eq:Hamil-total}
\end{equation}
The Hamiltonian for each FM layer reads
\begin{eqnarray}
H_{n}&=&
-A_{\text{intra}}\sum_{\langle i,j\rangle}\boldsymbol{m}_{i}^{n}\cdot\boldsymbol{m}_{j}^{n}
+D_{ij}\sum_{\langle i,j\rangle}(\boldsymbol{\nu}_{ij}\times\hat{z})\cdot(\boldsymbol{m}_{i}^{n}\times\boldsymbol{m}_{j}^{n}) \nonumber \\
&&+K\sum_{i}[1-(m_{i}^{n,z})^{2}]+H_{\text{DDI}},
\label{eq:Hamil-intralayer}
\end{eqnarray}
where $n$ is the FM layer index ($n=1,2,\cdots,N$), $\boldsymbol{m}_{i}^{n}$ represents the local magnetic moment orientation normalized as $|\boldsymbol{m}_{i}^{n}|=1$ at the site $i$, and $\left\langle i,j\right\rangle$ runs over all the nearest-neighbor sites in each FM layer. The first term represents the intralayer FM exchange interaction with the intralayer FM exchange stiffness $A_{\text{intra}}$. The second term represents the DMI with the DMI coupling energy $D_{ij}$, where $\boldsymbol{\nu}_{ij}$ is the unit vector between sites $i$ and $j$. The third term represents the PMA with the anisotropy constant $K$. $H_{\text{DDI}}$ represents the dipole-dipole interaction. When $N>1$, there exists an AFM exchange coupling between the nearest-neighbor FM layers
\begin{equation}
H_{\text{inter}}=-\sum_{n=1}^{N-1}A_{\text{inter}}\sum_{i}\boldsymbol{m}_{i}^{n}\cdot\boldsymbol{m}_{i}^{n+1}.
\label{eq:Hamil-interlayer}
\end{equation}
The sign of the interlayer exchange stiffness $A_{\text{inter}}$ is negative for the interlayer AFM exchange interaction. We take the initial magnetization direction in the FM layer L1 to be pointing upward (see Fig.~\ref{FIG1}).

\begin{figure}[t]
\centerline{\includegraphics[width=0.50\textwidth]{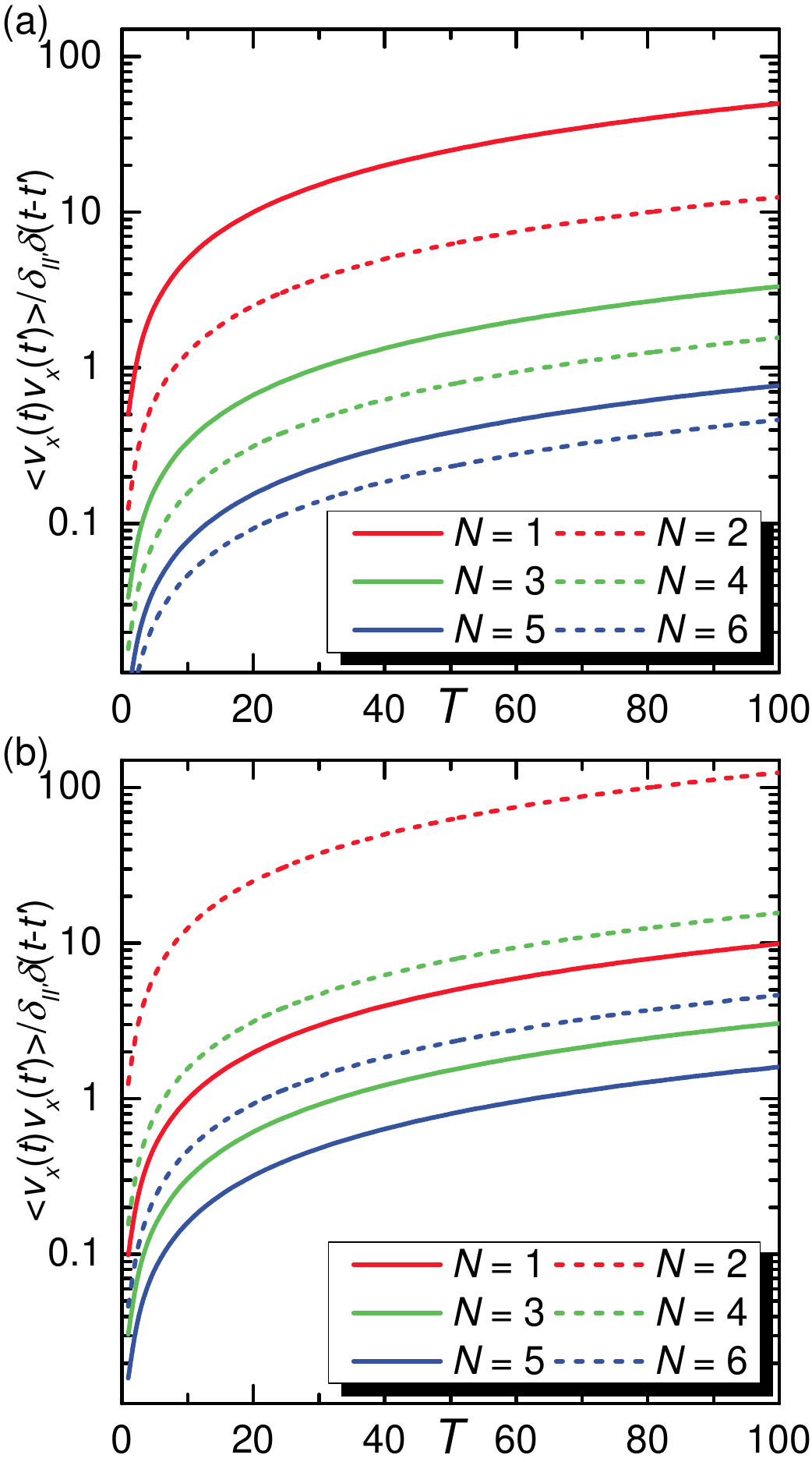}}
\caption{(Color online)
The velocity-velocity correlation function Eq.~(\ref{eq:v-v-correlations}) as a function of $T$ for the SAF $N$-layer skyrmion at (a) $\alpha=\D$ and (b) $\alpha=\D/10$. The solid curve shows the case of the SAF $N$-layer skyrmion with an odd $N$, which has $Q_{\text{tot}}=1$. The dashed curve shows the case of the SAF $N$-layer skyrmion with an even $N$, which has $Q_{\text{tot}}=0$. Here we assume $2k_{\text{B}}a^{2}/\hbar=\mathcal{D}=1$ in Eq.~(\ref{eq:v-v-correlations}).
}
\label{FIG2}
\end{figure}

\subsection{LLG equation at finite temperature}
\label{se:LLG}

The dynamics of a skyrmion at a given finite temperature is described by introducing a Gaussian stochastic magnetic field $\boldsymbol{h}$ describing the thermal agitation of the magnetization~\cite{Brown,Kubo,Duine,Mochizuki,Tronco}, which satisfies
\begin{equation}
\langle h_{i}(\boldsymbol{x},t)h_{j}(\boldsymbol{x}^{\prime},t^{\prime})\rangle=\frac{2\alpha k_{\text{B}}T}{\hbar}a^{2}\delta(\boldsymbol{x}-\boldsymbol{x}^{\prime})\delta_{ij}\delta(t-t^{\prime}),
\label{eq:h}
\end{equation}
where $i,j=x,y$, and $a^{2}$ is the area of the lattice. In the CPP geometry, we numerically solve the Landau-Lifshitz-Gilbert (LLG) equation including the spin-transfer torque (STT) term extended into the following form
\begin{align}
\frac{d\boldsymbol{m}_{i}}{dt}=&-|\gamma|\boldsymbol{m}_{i}\times(\boldsymbol{H}_{i}^{\text{eff}}+\boldsymbol{h})+\alpha\boldsymbol{m}_{i}\times\frac{d\boldsymbol{m}_{i}}{dt} \notag \\
&+\left\vert\gamma\right\vert u(\boldsymbol{m}_{i}\times\boldsymbol{p}\times\boldsymbol{m}_{i}),
\label{eq:LLGS}
\end{align}
with the layer index $n$ suppressed. Here, $\boldsymbol{H}_{i}^{\text{eff}}=-\partial H_{\text{total}}/\partial\mathbf{m}_{i}$ is the effective magnetic field induced by the total Hamiltonian $H_{\text{total}}$, $\gamma$ is the gyromagnetic ratio, $\alpha$ is the Gilbert damping coefficient originating from the spin relaxation, $u$ is the STT coefficient, and $\boldsymbol{p}$ represents the unit spin polarization vector of the spin current. We have $u=|\frac{\hbar}{\mu_{0}e}|\frac{jP}{2dM_{\text{S}}}$ with $\mu_{0}$ the vacuum magnetic permittivity, $d$ the thickness of the FM layer, $M_{\text{S}}$ the saturation magnetization, $j$ the applied current density, and $P$ the spin polarization rate. The STT exerted on the FM layer L1 is induced by the spin Hall effect. It should be noted that we have $j=0$ in the FM layers above the FM layer L1, thus there is no STT effect on the FM layer with the layer index number $n>1$.

\subsection{SkHE in multilayer SAF racetracks}
\label{se:SkHE}

\begin{figure*}[t]
\centerline{\includegraphics[width=1.00\textwidth]{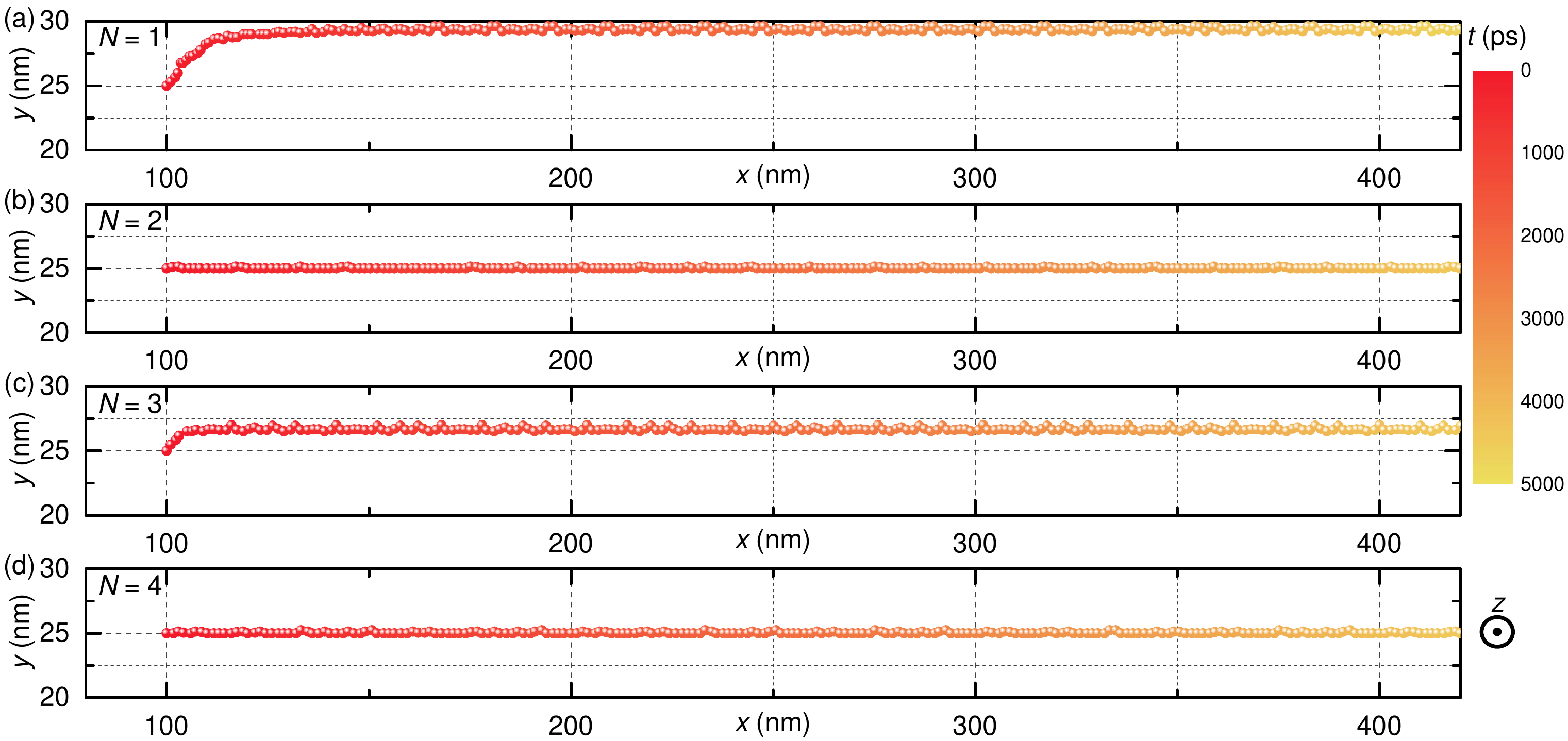}}
\caption{(Color online)
Typical trajectories of SAF $N$-layer skyrmions in $N$-layer SAF racetracks.
(a) The trajectory of a FM monolayer skyrmion in a monolayer FM racetrack ($N=1$) at $j=10$ MA cm$^{-2}$. The transverse shift of the monolayer skyrmion due to the SkHE is obvious. It reaches a stable velocity of $v_x\sim 70$ m s$^{-1}$.
(b) The trajectory of a SAF bilayer skyrmion in a bilayer SAF racetrack ($N=2$) at $j=20$ MA cm$^{-2}$. The SAF bilayer skyrmion moves along the central line ($y=25$ nm) of the racetrack, which reaches a stable velocity of $v_x\sim 70$ m s$^{-1}$.
(c) The trajectory of a SAF trilayer skyrmion in a trilayer SAF racetrack ($N=3$) at $j=30$ MA cm$^{-2}$. It reaches a stable velocity of $v_x\sim 70$ m s$^{-1}$.
(d) The trajectory of a SAF quadrilayer skyrmion in a quadrilayer SAF racetrack ($N=4$) at $j=40$ MA cm$^{-2}$. The SAF quadrilayer skyrmion moves along the central line ($y=25$ nm) of the racetrack, which reaches a stable velocity of $v_x\sim 70$ m s$^{-1}$. The dot denotes the center of the skyrmion. The total simulation time is $5000$ ps, which is indicated by the color scale.
}
\label{FIG3}
\end{figure*}

\begin{figure*}[t]
\centerline{\includegraphics[width=1.00\textwidth]{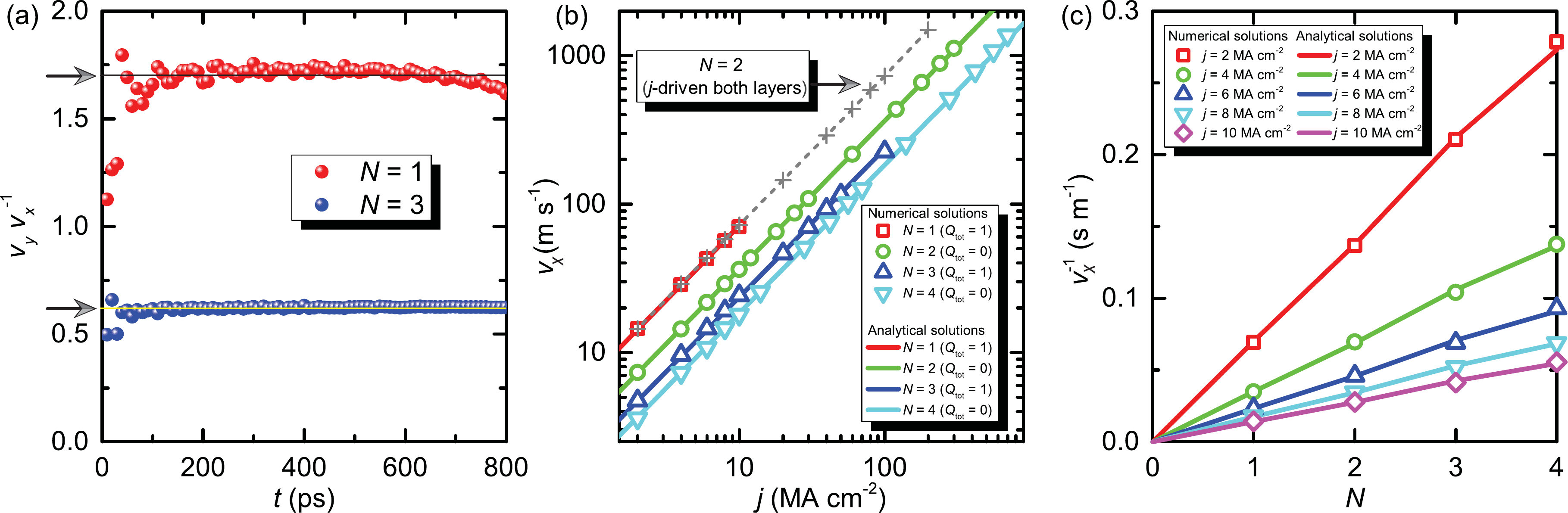}}
\caption{(Color online)
(a) The skyrmion Hall angle $v_{y}/v_{x}$ as a function of $t$. It is almost a constant for $N=1$ and $N=3$. Here, we use a square sample of $200$ nm $\times$ $200$ nm $\times$ $1$ nm in order to reduce the impact of the edge effect. The skyrmion is located at the film center ($100$ nm, $100$ nm) at the initial time.
(b) The velocity $v_{x}$ as a function of $j$ for the motion of SAF $N$-layer skyrmions, where the driving current is injected into the bottom FM layer L1. When $j>10$ MA cm$^{-2}$ and $j>100$ MA cm$^{-2}$, the moving FM monolayer and SAF trilayer skyrmions are destroyed en-route caused by the SkHE, respectively. The open symbols stand for the numerical results. The solid lines represent the theoretical results given by Eq.~(\protect\ref{eq:Mean-Velocity}) with $\protect\alpha\mathcal{D}=0.57$. The dashed line with cross symbol indicates the velocity $v_x$ of a SAF bilayer skyrmion in a bilayer SAF racetrack, where the driving current is injected into both the bottom FM layer L1 and the top FM layer L2.
(c) The inverse velocity $v_{x}^{-1}$ as a function of the total FM layer number $N$ at small $j$, where no skyrmion is destroyed by the SkHE. The open symbols stand for the numerical results. The solid lines represent the theoretical results given by Eq.~(\protect\ref{eq:Mean-Velocity}).
}
\label{FIG4}
\end{figure*}

We employ the Thiele equation~\cite{Thiele_PRL1973,Tomasello_SREP2014} with the inclusion of the stochastic force~\cite{Tronco} in order to interpret the numerical results. We generalize it to the SAF $N$-layer skyrmion system in multilayer SAF racetracks driven by the spin current with the CPP geometry at finite temperature. It would read in each FM layer as
\begin{equation}
\mathbf{G}_{n}\times\mathbf{v}^{n}-{\mathcal{D}}\alpha\mathbf{v}^{n}+\mathbf{j}_{\text{spin}}^{n}+\mathbf{I}_{\text{AFM}}^{n}=\mathbf{\eta}^{n},
\label{eq:ThieleEq-layer}
\end{equation}
with $n$ the layer index, where $\mathbf{v}^{n}$, $\mathbf{j}_{\text{spin}}^{n}$ and $\mathbf{I}_{\text{AFM}}^{n}$ represent the skyrmion velocity, the spin current, and the interlayer AFM exchange force, respectively. $\mathbf{G}_{n}=(0,0,4\pi Q_{n})$ is the gyromagnetic coupling constant representing the Magnus force with $Q_{n}$ the skyrmion number, which is defined as
\begin{equation}
Q_{n}=-\frac{1}{4\pi}\int\boldsymbol{m}^{n}(\boldsymbol{x})\cdot\left(\partial_{x}\boldsymbol{m}^{n}(\boldsymbol{x})\times\partial_{y}\boldsymbol{m}^{n}(\boldsymbol{x})\right)d^{2}\boldsymbol{x},
\label{eq:SkNum}
\end{equation}
and $\mathbf{\eta}^{n}$ is the Gaussian stochastic forces acting on the skyrmions representing the finite temperature effect, which satisfies
\begin{equation}
\langle\eta_{i}^{n}(t)\eta_{j}^{n}(t^{\prime})\rangle=\frac{2\alpha k_{\text{B}}T}{\hbar}a^{2}\mathcal{D}\delta_{ij}\delta(t-t^{\prime}),
\label{eq:Eta}
\end{equation}
where $i,j=x_{n},y_{n}$. We have taken the same dissipation matrix ${\mathcal{D}}$ and the same damping coefficient $\alpha$ for all racetracks.

We now postulate that all skyrmions move together with the same velocity $\mathbf{v}$ since they are tightly bound. Summing all $N$ Thiele Eqs.~(\ref{eq:ThieleEq-layer}), we would phenomenologically obtain
\begin{equation}
\mathbf{G}_{\text{tot}}\times\mathbf{v}-N{\mathcal{D}}\alpha\mathbf{v}+\mathbf{j}_{\text{spin}}=\mathbf{\eta}_{\text{tot}},
\label{eq:ThieleEq-total}
\end{equation}
where the interlayer AFM forces are assumed to be canceled out, that is, $\sum\mathbf{I}_{\text{AFM}}^{n}=0$, and $\mathbf{G}_{\text{tot}}=(0,0,4\pi Q_{\text{tot}})$ with
\begin{equation}
Q_{\text{tot}}=\sum_{n=1}^{N}Q_{n},
\label{eq:SkNum-total}
\end{equation}
and $\mathbf{j}_{\text{spin}}=\sum_{n=1}^{N}\mathbf{j}_{\text{spin}}^{n}$, $\mathbf{\eta}_{\text{tot}}=\sum_{n=1}^{N}\mathbf{\eta}^{n}$. Actually, $\mathbf{j}_{\text{spin}}^{n}=\delta_{n1}\mathbf{j}_{\text{spin}}$, where $\mathbf{j}_{\text{spin}}$ is the spin current induced by the charge current in the heavy-metal substrate due to the spin Hall effect (see Sec.~\ref{se:Monolayer-and-multilayer}).

The first term on the left hand side of Eq.~(\ref{eq:ThieleEq-total}) corresponds to the Magnus force. The total skyrmion number equals one, $Q_{\text{tot}}=1$, when the number $N$ of the FM layers is odd, while it equals zero, $Q_{\text{tot}}=0$, when the number $N$ of the FM layers is even since $Q_{n}=-(-1)^{n}$. The variant of the sum of the Gaussian noise is given by
\begin{equation}
\langle\eta_{\text{tot}}^{i}(t)\eta_{\text{tot}}^{j}(t^{\prime})\rangle=\frac{2\alpha k_{\text{B}}T}{N\hbar}a^{2}\mathcal{D}\delta_{ij}\delta(t-t^{\prime}),
\label{eq:Eta-total}
\end{equation}
where $i,j=x,y$. As a result, the variance of the total noise becomes $1/N$ in the SAF $N$-layer skyrmion. Therefore, the SAF $N$-layer skyrmion is $N$ times more stable than the FM monolayer skyrmion at a certain temperature and a certain damping coefficient.

The velocity is given by explicitly solving the Thiele Eq.~(\ref{eq:ThieleEq-total}) as
\begin{widetext}
\begin{eqnarray}
v_{x}&=&\frac{\alpha N{\mathcal{D}}}{Q_{\text{tot}}^{2}
+{\alpha^{2}N}^{2}{{\mathcal{D}}^{2}}}\left(j_{\text{spin}}-\eta_{\text{tot}}^{x}\right)
+\frac{Q_{\text{tot}}}{Q_{\text{tot}}^{2}+{\alpha^{2}N}^{2}{{\mathcal{D}}^{2}}}\eta_{\text{tot}}^{y}, \quad \\
v_{y}&=&\frac{Q_{\text{tot}}}{Q_{\text{tot}}^{2}+{\alpha^{2}N}^{2}
{{\mathcal{D}}^{2}}}\left(j_{\text{spin}}-\eta_{\text{tot}}^{x}\right)
-\frac{\alpha N{\mathcal{D}}}{Q_{\text{tot}}^{2}+{\alpha^{2}N}^{2}{{\mathcal{D}}^{2}}}\eta_{\text{tot}}^{y}.
\label{eq:vx-vy}
\end{eqnarray}
The mean velocity is thus given by
\begin{equation}
\left\langle v_{x}\right\rangle=\frac{\alpha N{\mathcal{D}}}{Q_{\text{tot}}^{2}
+{\alpha^{2}N}^{2}{{\mathcal{D}}^{2}}}j_{\text{spin}},\quad
\left\langle v_{y}\right\rangle=\frac{Q_{\text{tot}}}{Q_{\text{tot}}^{2}
+{\alpha^{2}N}^{2}{{\mathcal{D}}^{2}}}j_{\text{spin}}.
\label{eq:Mean-Velocity}
\end{equation}
\end{widetext}
When $Q_{\text{tot}}=1$, which is the case for $N$ being odd, a skyrmion undergoes a transverse motion, where
\begin{equation}
\frac{\left\langle v_{y}\right\rangle}{\left\langle v_{x}\right\rangle}=\frac{1}{\alpha N{\mathcal{D}}}.
\label{eq:HallAngle}
\end{equation}
When $Q_{\text{tot}}=0$, which is the case for $N$ being even, a skyrmion goes straight, where
\begin{equation}
\left\langle v_{x}\right\rangle=\frac{1}{\alpha N{\mathcal{D}}}j_{\text{spin}}, \quad
\left\langle v_{y}\right\rangle=0.
\label{eq:vx-vy-Q-0}
\end{equation}
Consequently the SAF $N$-layer skyrmion experiences the SkHE only when $N$ is odd. We call it the odd-even effect on the SkHE.

The velocity of the skyrmion decreases with increasing $N$ at a given driving current density. The velocity-velocity correlation functions are calculated as
\begin{widetext}
\begin{eqnarray}
\langle v_{x}(t)v_{x}(t^{\prime })\rangle&=&\frac{\left(\alpha N{\mathcal{D}}\right)^{2}\langle\eta_{\text{tot}}^{x}(t)\eta_{\text{tot}}^{x}(t^{\prime})\rangle+Q_{\text{tot}}^{2}\langle\eta_{\text{tot}}^{y}(t)\eta_{\text{tot}}^{y}(t^{\prime})\rangle}{\left(Q_{\text{tot}}^{2}+{\alpha^{2}N}^{2}{{\mathcal{D}}^{2}}\right)^{2}}, \\
\langle v_{y}(t)v_{y}(t^{\prime})\rangle&=&\frac{Q_{\text{tot}}^{2}\langle\eta_{\text{tot}}^{x}(t)\eta_{\text{tot}}^{x}(t^{\prime})\rangle+\left(\alpha N{\mathcal{D}}\right)^{2}\langle\eta_{\text{tot}}^{y}(t)\eta_{\text{tot}}^{y}(t^{\prime})\rangle}{\left(Q_{\text{tot}}^{2}+{\alpha^{2}N}^{2}{{\mathcal{D}}^{2}}\right)^{2}}.
\label{eq:vx-vy-correlations}
\end{eqnarray}
Substituting Eq.~(\ref{eq:Eta-total}), the correlation functions are obtained as
\begin{eqnarray}
\langle v_{x}(t)v_{x}(t^{\prime })\rangle&=&\langle v_{y}(t)v_{y}(t^{\prime })\rangle \nonumber \\
&=&\frac{1}{Q_{\text{tot}}^{2}+{\alpha^{2}N}^{2}{{\mathcal{D}}^{2}}}\frac{2\alpha k_{\text{B}}T}{N\hbar}a^{2}\mathcal{D}\delta_{II^{\prime}}\delta(t-t^{\prime}),
\label{eq:v-v-correlations}
\end{eqnarray}
\end{widetext}
which are functions of $T$ and $\alpha$ for a given SAF $N$-layer skyrmion.

Since the scope of this paper is focused on the temperature effect on the SAF $N$-layer skyrmion, we assume $2k_{\text{B}}a^{2}/\hbar=\mathcal{D}=1$ in Eq.~(\ref{eq:v-v-correlations}) and show the correlation functions as a function of $T$ for the SAF $N$-layer skyrmions under the assumptions of a large damping coefficient ($\alpha=\D$) and a small damping coefficient ($\alpha=\D/10$) in Figs.~\ref{FIG2}(a) and \ref{FIG2}(b), respectively.

It can be seen that, for the case of large damping coefficient ($\alpha=\D$) [Fig.~\ref{FIG2}(a)], the correlation functions of the $N$-layer skyrmion, where $N=1,2,\cdots,6$, increase with increasing $T$. On the other hand, the correlation functions are inversely proportional to $N$. This indicates that the stability of the SAF $N$-layer skyrmion with a large damping coefficient increases with $N$ but decreases with $T$. In contrast, for the case of small damping coefficient ($\alpha=\D/10$) [Fig.~\ref{FIG2}(b)], it shows the correlation functions of the SAF $N$-layer skyrmion, where $N=1,2,\cdots,6$, increase with increasing $T$. However, the correlation functions are nonmonotonic with respect to $N$. At a certain $T$, it can be seen that the correlation functions of the SAF $N$-layer skyrmions with $N=2,4$ are larger than that of the SAF $N$-layer skyrmion with $N=1$. This means that the stability of the SAF $N$-layer skyrmion with a small damping coefficient decreases with $T$, however, the FM monolayer skyrmion with $Q_{\text{tot}}=1$ is more stable than the SAF bilayer skyrmion with $Q_{\text{tot}}=0$ at a certain $T$. It is noteworthy that this result is in good agreement with a recent study on the AFM skyrmion (see Ref.~\onlinecite{Tretiakov_PRL2016}), where the fluctuation of the AFM skyrmion with $Q_{\text{tot}}=0$ is found to be inversely proportional to $\alpha$, and is more significant than that of the FM skyrmion with $Q_{\text{tot}}=1$ at a small damping coefficient $\alpha=0.01$. It is also worth mentioning that the fluctuation of the FM skyrmion with $Q_{\text{tot}}=1$ is proportional to $\alpha$ (see Refs.~\onlinecite{Tretiakov_PRL2016,Schutte_PRB2014}). Hence, in order to remain in a monotonic temperature dependence with respect to $N$, we numerically study the SAF $N$-layer skyrmion under the large damping coefficient assumption in the following sections.

\begin{figure*}[t]
\centerline{\includegraphics[width=1.00\textwidth]{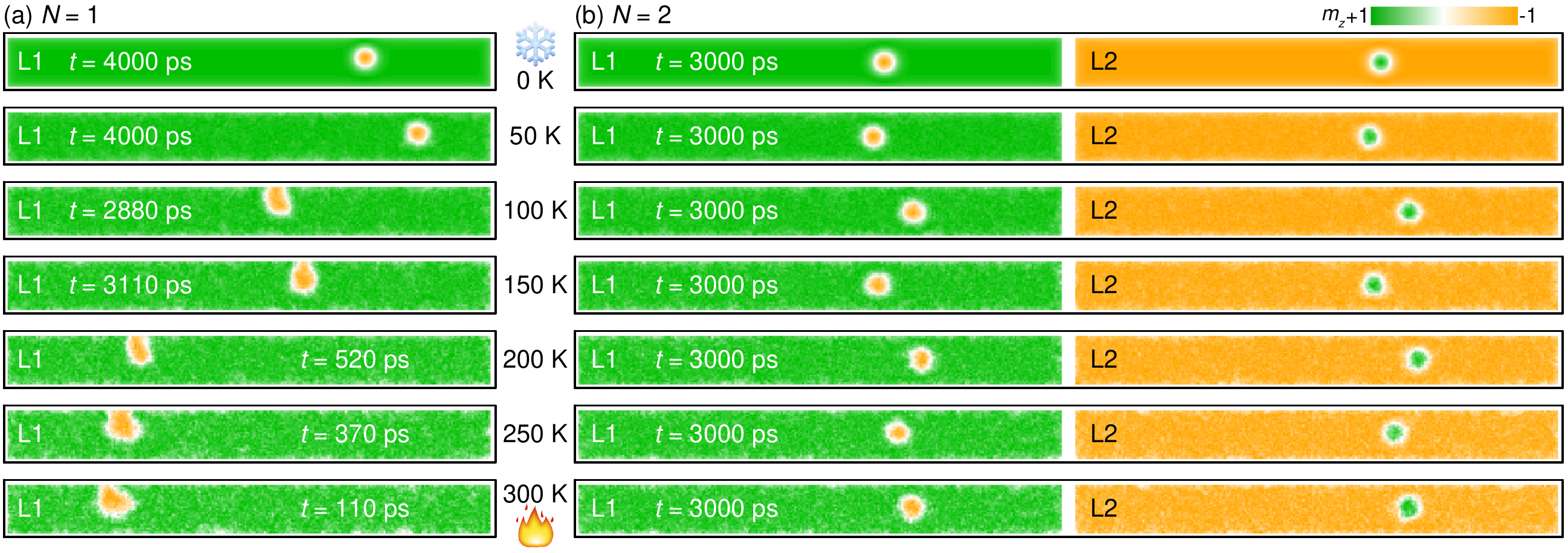}}
\caption{(Color online)
Top views of the motion of a FM monolayer skyrmion and a SAF bilayer skyrmion at different $T$ and selected $t$.
(a) The motion of a FM monolayer skyrmion ($Q_{\text{tot}}=1$) in a monolayer FM racetrack ($N=1$). A moderate driving current of $j=10$ MA cm$^{-2}$ is applied. The FM monolayer skyrmion driven by the spin current moves safely from the left to the right end of the racetrack at $T=0$ and $50$ K. However, when $T\geq 100$ K, it is destroyed by touching the upper edge.
(b) The motion of a SAF bilayer skyrmion ($Q_{\text{tot}}=0$) in a bilayer SAF racetrack ($N=2$). A moderate driving current of $j=20$ MA cm$^{-2}$ is applied. The SAF bilayer skyrmion, which is immune from the SkHE, reliably moves along the central line ($y=25$ nm) of the racetrack even at $T=300$ K. The seed of the random number generator used to generate the thermal fluctuation field is set to $100$ in all simulations with $T>0$ K. The average out-of-plane magnetization component $m_z$ is denoted by the green-white-orange color scale.
}
\label{FIG5}
\end{figure*}

\begin{figure*}[t]
\centerline{\includegraphics[width=1.00\textwidth]{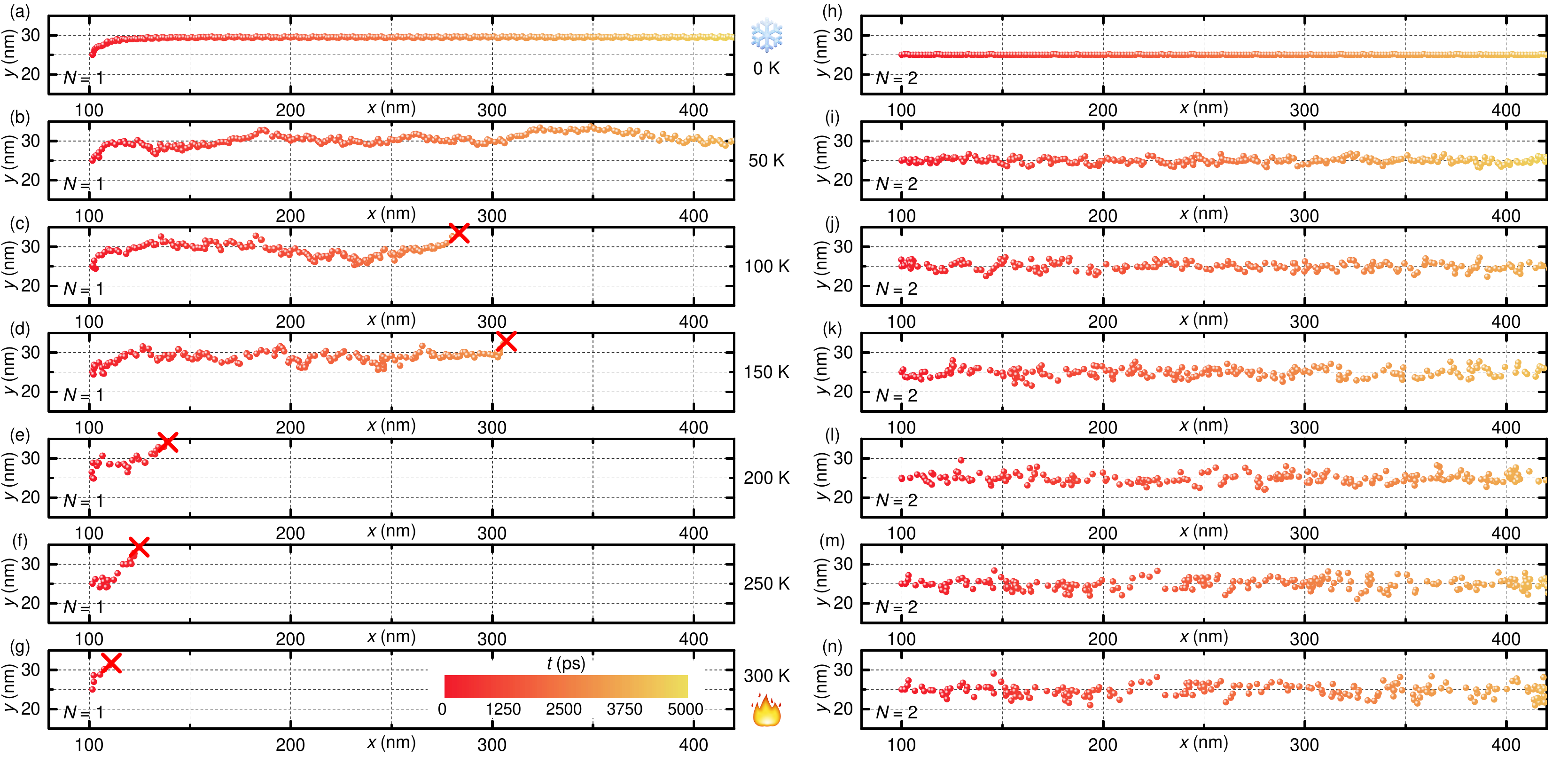}}
\caption{(Color online)
Typical trajectories of a FM monolayer skyrmion and a SAF bilayer skyrmion at different $T$. Trajectories of a FM monolayer skyrmion ($Q_{\text{tot}}=1$) in a monolayer FM racetrack at (a) $T=0$ K, (b) $T=50$ K, (c) $T=100$ K, (d) $T=150$ K, (e) $T=200$ K, (f) $T=250$ K, and (g) $T=300$ K. A moderate driving current of $j=10$ MA cm$^{-2}$ is applied. Trajectories of a SAF bilayer skyrmion ($Q_{\text{tot}}=0$) in a bilayer SAF racetrack at (h) $T=0$ K, (i) $T=50$ K, (j) $T=100$ K, (k) $T=150$ K, (l) $T=200$ K, (m) $T=250$ K, and (n) $T=300$ K. A moderate driving current of $j=20$ MA cm$^{-2}$ is applied. The seed of the random number generator used to generate the thermal fluctuation field is set to $100$ in all simulations with $T>0$ K. The total simulation time is $5000$ ps, which is represented by the color scale. The dot denotes the center of the skyrmion. The red cross indicates the destruction of the skyrmion by touching the upper edge caused by the SkHE.
}
\label{FIG6}
\end{figure*}

\begin{figure*}[t]
\centerline{\includegraphics[width=1.00\textwidth]{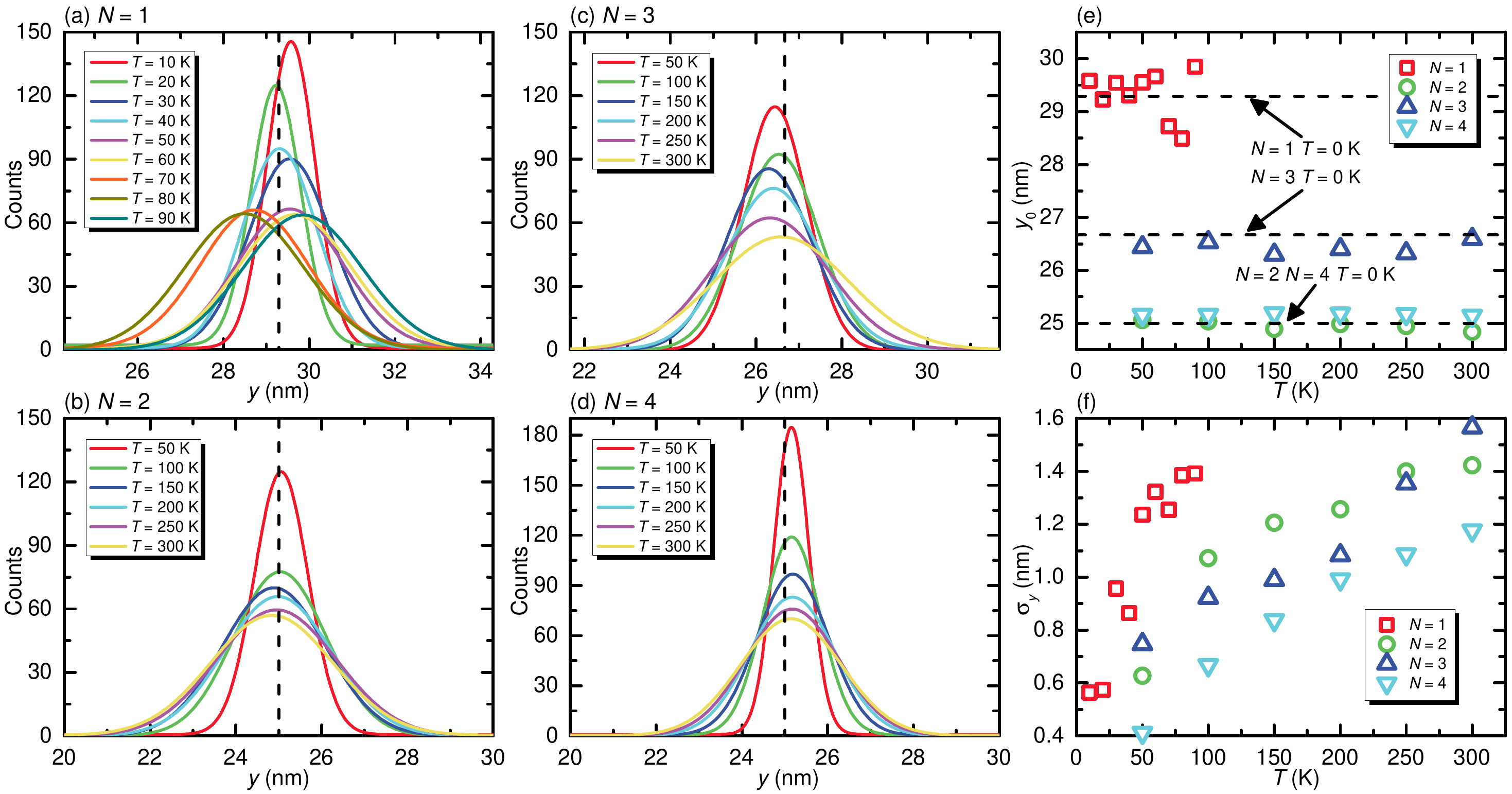}}
\caption{(Color online)
Distributions of the $y$-position of SAF $N$-layer skyrmions at different $T$ fitted by the Gaussian distribution. (a) A FM monolayer skyrmion ($Q_{\text{tot}}=1$) driven by a current of $j=10$ MA cm$^{-2}$. At $T=0$ K, the $y$-position equals $\sim 29.3$ nm. In order to ensure a safe motion of $\sim 300$ nm, $T$ is only increased up to $90$ K. (b) A SAF bilayer skyrmion ($Q_{\text{tot}}=0$) driven by a current of $j=20$ MA cm$^{-2}$. At $T=0$ K, the $y$-position equals $\sim 25.0$ nm. (c) A SAF trilayer skyrmion ($Q_{\text{tot}}=1$) driven by a current of $j=30$ MA cm$^{-2}$. At $T=0$ K, the $y$-position equals $\sim 26.7$ nm. (d) A SAF quadrilayer skyrmion ($Q_{\text{tot}}=0$) driven by a current of $j=40$ MA cm$^{-2}$. At $T=0$ K, the $y$-position equals $\sim 25.0$ nm. (e) The mean $y_0$ of the distribution of the $y$-position as a function of $T$. (f) The standard deviation $\protect\sigma_y$ of the distribution of the $y$-position as a function of $T$. The $y$-position of the skyrmion at $T=0$ K is indicated by the vertical dashed line in (a-d). The seed of the random number generator used to generate the thermal fluctuation field is set to $100$ in all simulations with $T>0$ K.
}
\label{FIG7}
\end{figure*}

\begin{figure*}[t]
\centerline{\includegraphics[width=1.00\textwidth]{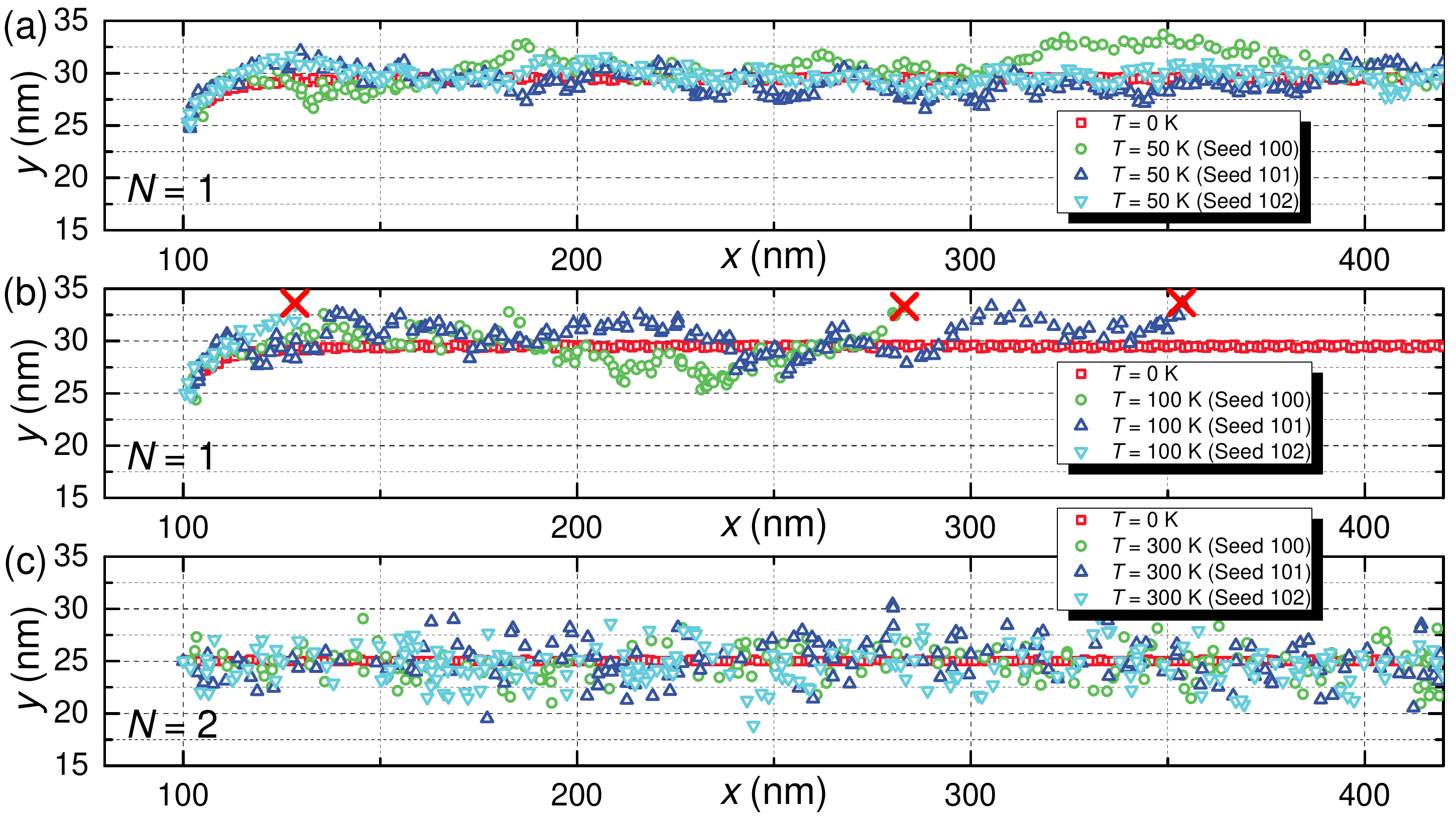}}
\caption{(Color online)
Trajectories of a FM monolayer skyrmion and a SAF bilayer skyrmion at selected $T$ with several different random seed values. Trajectories of a FM monolayer skyrmion ($Q_{\text{tot}}=1$) (a) at $T=0,50$ K and (b) at $T=0,100$ K with a driving current of $j=10$ MA cm$^{-2}$. (c) Trajectories of a SAF bilayer skyrmion ($Q_{\text{tot}}=0$) at $T=0, 300$ K with a driving current of $j=20$ MA cm$^{-2}$. The seed of the random number generator used to generate the thermal fluctuation field is set to $100$, $101$, and $102$, respectively. The total simulation time is $5000$ ps. The symbol denotes the center of the skyrmion. The red cross indicates the destruction of the skyrmion by touching the upper edge caused by the SkHE.
}
\label{FIG8}
\end{figure*}

\subsection{Simulation methods}
\label{se:Simulation-methods}

The three-dimensional micromagnetic simulations are performed by using the 1.2 alpha 5 release of the Object Oriented MicroMagnetic Framework (OOMMF) software developed at the National Institute of Standards and Technology (NIST)~\cite{OOMMF}. The simulations are handled by the OOMMF extensible solver (OXS) objects of the standard OOMMF distribution with the OXS extension modules for including the interface-induced DMI~\cite{OOMMFDMI,Rohart_PRB2013} and the thermal fluctuation~\cite{OOMMFXF_ThermSpinXferEvolve}. The magnetization dynamics at zero temperature ($T=0$ K) is controlled by the LLG equation including the STT term~\cite{OOMMF,LLGSTT}, while a highly irregular fluctuating field representing the irregular influence of temperature is added into the LLG equation including the STT term when the thermal effect is considered ($T>0$ K). The finite temperature simulations are performed with a fixed time step of $10$ fs, that is, $1\times 10^{-14}$ s, while the time step in the zero temperature simulations is adaptive ($\sim 1\times 10^{-13}$ s). Each simulation with a certain finite temperature is performed $10$ times individually with different random seed values. The models built in the micromagnetic simulations are discretized into regular cells with a constant cell size of $2$ nm $\times$ $2$ nm $\times$ $1$ nm, which allows for a trade-off between numerical accuracy and computational efficiency. In the CPP geometry, the spin current polarized along the $-y$-direction flows upward in the bottom FM layer L1, which is induced by the charge current flowing in the heavy-metal substrate due to the spin Hall effect. The current-induced Oersted field is neglected in all simulations for simplicity since it makes only a minor contribution to the overall magnetization dynamics. The typical material parameters used in the micromagnetic simulations are adopted from Refs.~\cite{Fert_NNANO2013,Sampaio_NNANO2013,Tomasello_SREP2014,Xichao_NCOMMS2016}: Gilbert damping coefficient $\alpha=0.3$ which is large enough to satisfy the large damping coefficient assumption; gyromagnetic ratio $\gamma=-2.211\times 10^{5}$ m A$^{-1}$ s$^{-1}$; saturation magnetization $M_{\text{S}}=580$ kA m$^{-1}$; intralayer FM exchange stiffness $A_{\text{intra}}=15$ pJ m$^{-1}$; interlayer AFM exchange stiffness $A_{\text{inter}}=-1$ pJ m$^{-1}$; interface AFM exchange coefficient $\sigma=-1$ mJ m$^{-2}$; DMI constant $D=3.5$ mJ m$^{-2}$; PMA constant $K=0.8$ MJ m$^{-3}$; and spin-polarization rate $P=0.4$.

\section{Current-induced motion of skyrmions at zero temperature}
\label{se:motion-at-zero-temperature}

We start with a numerical investigation of the current-velocity relation of skyrmions in $N$-layer SAF racetracks, with $N=1,2,3,4$, at zero temperature ($T=0$ K) (see Fig.~\ref{FIG3}).

Let us recapitulate the current-induced motion of a FM monolayer skyrmion in a monolayer FM racetrack, which undergoes a transverse motion toward the upper edge of the racetrack because of the SkHE (see Ref.~\onlinecite{SI} for Movie 1). We show the trajectory at a moderate driving current of $j=10$ MA cm$^{-2}$ in Fig.~\ref{FIG3}(a). The moving skyrmion reaches a stable velocity of $v_x\sim 70$ m s$^{-1}$ and has a transverse shift of $\sim 5$ nm due to the SkHE. It does not touch the edge because of the repulsive force from the edge.
Nevertheless, when $j>10$ MA cm$^{-2}$, it is destroyed by touching the edge shortly after the driving current is applied (see Ref.~\onlinecite{SI} for Movie 1).

Let us also recapitulate the current-induced motion of a SAF bilayer skyrmion in a bilayer SAF racetrack (see Ref.~\onlinecite{SI} for Movie 2). It goes straight in the bilayer SAF racetrack as a result of the suppression of the SkHE ($Q_{\text{tot}}=0$). The trajectory at a moderate driving current of $j=20$ MA cm$^{-2}$ is shown in Fig.~\ref{FIG3}(b), where it reaches a stable speed of $\sim 70$ m s$^{-1}$. The SAF bilayer skyrmion strictly moves along the central line ($y=25$ nm) of the racetrack.

We go on to study the current-induced motion of a SAF trilayer skyrmion with $Q_{\text{tot}}=1$ [see Fig.~\ref{FIG3}(c)], and a SAF quadrilayer skyrmion with $Q_{\text{tot}}=0$ [see Fig.~\ref{FIG3}(d)]. The SAF trilayer skyrmion experiences the SkHE (see Ref.~\onlinecite{SI} for Movie 3) as in the case of the FM monolayer skyrmion. On the other hand, the SAF quadrilayer skyrmion moves reliably in the quadrilayer SAF racetrack even at a strong driving current (see Ref.~\onlinecite{SI} for Movie 4), demonstrating the suppression of the SkHE as in the case of a SAF bilayer skyrmion. The SAF quadrilayer skyrmion moves along the central line ($y=25$ nm) of the racetrack.

We compare the SkHEs at $N=1$ and $N=3$ quantitatively. We show the skyrmion Hall angle $v_{y}/v_{x}$ as a function of time $t$ in Fig.~\ref{FIG4}(a) for $N=1,3$. The skyrmion Hall angle is antiproportional to $N$. Namely, the SkHE for $N=3$ is three times smaller than that for $N=1$. The theoretical expectation Eq.~(\ref{eq:HallAngle}) explains the numerical data remarkably well with the choice of $\alpha\mathcal{D}=0.57$.

In Fig.~\ref{FIG4}(b), we show the velocity $v_{x}$ as a function of the applied driving current density $j$ for the motion of SAF $N$-layer skyrmions, where $N=1,2,3,4$. We have fitted the data successfully by theoretical expectation Eq.~(\ref{eq:Mean-Velocity}) with the use of $\alpha\mathcal{D}=0.57$. The velocity $v_{x}$ is almost antiproportional to $N$ as shown in Fig.~\ref{FIG4}(c). This is because the driving current is only applied to the bottom FM layer L1.

In order to further improve the $j$-$v$ relation of the SAF $N$-layer skyrmions, here taking the SAF bilayer skyrmion as an example, we also investigate the $j$-$v$ curve when the driving current is applied in both constituent FM layers of the bilayer SAF racetrack, which is plotted in Fig.~\ref{FIG4}(b) as a dashed curve. It can be seen that, when both FM layers are driven by the current, the $j$-$v$ relation of the SAF bilayer skyrmion matches well with that of a FM monolayer skyrmion moving in a monolayer FM racetrack at the small driving current regime. When $j=10$ MA cm$^{-2}$, the velocity of the FM monolayer skyrmion is $v_x=70$ m s$^{-1}$, while the velocity of the SAF bilayer skyrmion is $v_x=72$ m s$^{-1}$.

\section{Current-induced motion of skyrmions at finite temperature}
\label{se:motion-at-finite-temperature}

We proceed to investigate the effect of random thermal perturbations on the motion of SAF $N$-layer skyrmions. Figure~\ref{FIG5} demonstrates the motion of a FM monolayer (SAF bilayer) skyrmion driven by a moderate current of $j=10$ ($20$) MA cm$^{-2}$ at temperatures ranging from $T=0$ K to $T=300$ K (see Ref.~\onlinecite{SI} for Movies 5 and 6).

A FM monolayer skyrmion moves safely from the left to the right terminal of the racetrack at $T=0$ and $50$ K. However, when $T$ is increased and larger than $100$ K, it becomes unstable and is destroyed by touching the upper edge. Figures~\ref{FIG6}(a)-(g) show its trajectories at $T=0$, $50$, $100$, $150$, $200$, $250$, and $300$ K, respectively. At $T=0$ K, as we have already stated, the transverse shift is constant due to the balance between the SkHE and the edge effect. At a finite temperature, the motion is affected by thermal effect, where the transverse shift is fluctuating. It is destroyed when the fluctuated skyrmion touches the edge. Indeed, it is destroyed after having moved $\sim 200$ nm at $T=100$ K. When $T$ is raised to room temperature, that is, $T=300$ K, the FM monolayer skyrmion is destroyed very shortly after the driving current is applied.

On the other hand, a SAF bilayer skyrmion moves safely from the left to the right terminal of the racetrack even at $T=300$ K. The corresponding trajectories of the SAF bilayer skyrmion at $T=0$, $50$, $100$, $150$, $200$, $250$, and $300$ K, are given in Figs.~\ref{FIG6}(h)-(n), respectively. Since the SAF bilayer skyrmion exhibits no transverse shift caused by the SkHE, it is not destroyed by touching the edge as the racetrack width is wide enough here.

In Figs.~\ref{FIG7}(a)-(d), we show the distributions of the $y$-position of a SAF $N$-layer skyrmion for $\sim 300$ nm of motion at different $T$, which are fitted by using the Gaussian distribution. In Figs.~\ref{FIG7}(e)-(f), we show the mean value $y_{0}$ and standard deviation $\sigma_{y}$ of the distribution of the $y$-position as functions of $T$ corresponding to Figs.~\ref{FIG7}(a)-(d). The mean $y_{0}$ is around the one half of the sample width for $N=2$ and $N=4$ since the SAF $N$-layer skyrmion with an even $N$ goes straight reflecting the absence of the SkHE. On the other hand, the mean $y_{0}$ is away from the one half of the sample width for $N=1$ and $N=3$ due to the SkHE. The deviation is larger for $N=1$ than $N=3$, which indicates that the SkHE exerted on a FM monolayer skyrmion is larger than that on a SAF trilayer skyrmion.

In Fig.~\ref{FIG6} we see that a FM monolayer skyrmion can go further at $T=150$ K than at $T=100$ K. This occurs accidentally due to the choice of the thermal random seed, which is a number employed by the random number generator to generate the thermal fluctuation field in thermal simulations~\cite{OOMMFXF_ThermSpinXferEvolve}. In order to evaluate and examine the effect of the thermal random seed on the simulation results, we perform each simulation at a given $T$ $10$ times individually with different thermal random seed values. The trajectories of a FM monolayer skyrmion in a monolayer FM racetrack at $T=50$ and $100$ K with three selected random seed values are shown in Figs.~\ref{FIG8}(a)-(b), respectively, where a moderate driving current of $j=10$ MA cm$^{-2}$ is applied. It can be seen that the trajectories at a certain $T$ with different random seed values are modestly influenced by thermal perturbations. At $T=50$ K, although the motion of the FM monolayer skyrmion is fluctuating, the FM monolayer skyrmion reaches the right terminal of the racetrack in all $10$ simulations with different random seed values. At $T=100$ K, the FM monolayer skyrmion is destroyed by touching the upper edge of the racetrack in all $10$ simulations with different random seed values. On the other hand, the SAF bilayer skyrmion is safely conveyed between the two terminals without touching the edge in all these simulations even at $T=300$ K, as shown in Fig.~\ref{FIG8}(c).

\section{Conclusions}
\label{se:Conclusions}

We have studied the motion of skyrmions in multilayer SAF racetracks in contrast to the motion of skyrmions in conventional monolayer FM racetracks. The thermal effect on the motion of skyrmions in monolayer FM racetracks as well as multilayer SAF racetracks have been investigated by including random thermal perturbations in the micromagnetic simulations. We have found that a moving SAF bilayer skyrmion is much more stable than a moving FM monolayer skyrmion, even when the temperature effect is taken into account, since the two skyrmions consisting the SAF bilayer skyrmion are tightly bound by the interlayer AFM exchange coupling, and thus the SAF bilayer skyrmion is immune from the SkHE. Besides, we have shown that the detrimental effect of the SkHE on the moving FM monolayer skyrmion is enhanced as temperature increases, while the SAF bilayer skyrmion can safely move along the racetrack even at room temperature. In addition, the odd-even effect of the constituent FM layer number on the SkHE in multilayer SAF racetracks has also been demonstrated. Due to the suppression of the SkHE, the skyrmions have no transverse motion in multilayer SAF racetracks with even constituent FM layers. In conclusion, we find that the bilayer SAF racetrack is a preferred host for skyrmion transmission in racetrack-type device applications since it realizes a minimum system which does not show the SkHE.

\begin{acknowledgments}
X.Z. was supported by the RONPAKU program of the Japan Society for the Promotion of Science. M.E. acknowledges the support by the Grants-in-Aid for Scientific Research from JSPS KAKENHI (Grants No. 25400317 and No. 15H05854). Y.Z. acknowledges the support by National Natural Science Foundation of China (Project No. 1157040329), the Seed Funding Program for Basic Research and Seed Funding Program for Applied Research from the HKU, ITF Tier 3 funding (ITS/203/14), the RGC-GRF under Grant HKU 17210014, and University Grants Committee of Hong Kong (Contract No. AoE/P-04/08). X.Z. thanks X.X. Liu for getting him involved in the JSPS RONPAKU program. M.E. is very much grateful to N. Nagaosa for many helpful discussions on the subject.
\end{acknowledgments}



\end{document}